\documentclass[aps,twocolumn,groupedaddress]{revtex4}

\begin{document}
\draft

\title{  Mesoscopic Transport of Entangled and Nonentangled Kondo Singlets under Bias}
\author{Jongbae Hong}
\address{Department of Physics,
Pohang University of Science and Technology, Pohang 790-784, Korea
\\ \& Asia Pacific Center for Theoretical Physics, Pohang, Gyeongbuk 790-784, Korea}
\date{\today}

\begin{abstract}

The unexplained tunneling conductances of correlated mesoscopic Kondo systems
are understood by the coherent transport of the entangled and nonentangled singlets.
Spins of the entangled singlet flow unidirectionally in a sequential up-and-down manner.
This dynamics does not follow linear response theory. The side
peaks at a finite bias are formed by resonant tunneling of the nonentangled singlet
through a coherent tunneling level formed by two electron reservoirs within a coherent region.
The theoretical line shapes remarkably fit the experimental data of a quantum point
contact and a magnetized atom adsorbed on an insulating layer covering metallic substrate.

\end{abstract}


\maketitle \narrowtext



\begin{figure}
[b] \vspace*{1.5cm} \includegraphics{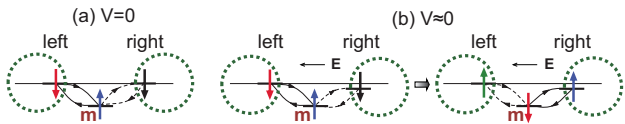} \vspace*{0.0cm}
\caption{(Color online) (a) An entangled Kondo singlet at $V=0$.
The dashed circle and the letter ``m" denote the Kondo cloud and
the mediating Kondo atom, respectively. (b) An entangled Kondo
singlet at $V\approx 0$. Singlet hopping and partner changing
leads to unidirectional spin flow. The horizontal arrow denotes
the direction of the electric field.
 }
\end{figure}

The nonlinear line shapes of the tunneling conductance observed for mesoscopic Kondo systems
are waiting for relevant theoretical explanations from the microscopic point of view.
The two-reservoir Anderson impurity model at steady-state nonequilibrium
is considered as a proper microscopic model describing
a mesoscopic Kondo system. However, previous theoretical studies using the noncrossing
approximation~\cite{wingreen2}, the Keldysh formalism~\cite{fujii}, quantum Monte Carlo
calculations~\cite{han}, and the extended numerical renormalization group method~\cite{anders}
do not reproduce various nonlinear line shapes of mesoscopic Kondo systems.
The difficulty originates from the combination of strong correlation and steady-state
nonequilibrium. To solve this problem, we need to understand the dynamics under bias more clearly.
It is obvious that the two-reservoir mesoscopic Kondo system at equilibrium has an entangled Kondo
singlet represented by a wave function
$|\Psi\rangle_{en}=p\!\!\mid\downarrow\uparrow\rangle_{LS}+q\!\!\mid\uparrow\downarrow\rangle_{LS}+
u\!\!\mid\downarrow\uparrow\rangle_{RS}+v\!\!\mid\uparrow\downarrow\rangle_{RS}$,
where $LS$ and $RS$ denote left singlet and right singlet,
respectively and the coefficients are complex numbers, as shown in
Fig.~1 (a). The equilibrium dynamics comprises multiple processes
of exchange, partner change, and singlet hopping in a mixed manner
among the four states given above. This complicated dynamics may
be studied using the numerical renormalization group method for
the low-energy regime~\cite{wilson}. Under bias, however, the
dynamics becomes considerably simpler because the spins involved
in the entanglement flow unidirectionally in an up-and-down
sequence, as shown in Fig.~1 (b), until the entanglement is
retained. The processes of singlet hopping and partner changing
are used in the spin flow, and the coherent spins are provided
from the Kondo cloud. The unidirectional flow does not allow a
linear response regime. Backward motion of electrons may occur in
the incoherent dynamics, and this leads to double occupancy at the
mediating atom.

In this study, we clarify the spin dynamics forming the zero-bias peak and
the side peaks and compare the theoretical results with the experimental data
obtained for a quantum point contact~\cite{sarkozy} and an
adsorbed magnetized atom on an insulating layer covering a metallic substrate~\cite{otte},
for example. Fitting the entire range of the line shapes of these systems
is given for the first time.

\begin{figure}
[b] \vspace*{2.5cm} \includegraphics{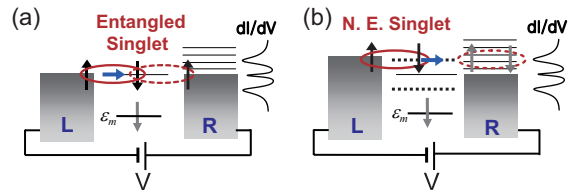} \vspace*{0.0cm}
\caption{ (Color online) (a) Transport of an entangled Kondo
singlet at low bias. (b) Resonant tunneling of a nonentangled (N. E.)
Kondo singlet through the coherent tunneling level (dashed line).
$\varepsilon_m$ denotes the energy level of the mediating spin
that forms a singlet. }
\end{figure}

In Fig.~2, we depict the low-energy tunneling schemes in the
two-reservoir Anderson impurity model under bias along with the
corresponding line shape of the tunneling conductance. Figure 2
(a) describes the transport near zero bias, i.e., $eV<k_BT_K$,
where $e$, $V$, $k_B$, and $T_K$ denote the electron charge,
source-drain bias, Boltzmann's constant, and Kondo temperature for
the entangled Kondo singlet, respectively. The unidirectional flow
described in Fig.~1 (b) establishes the zero-bias peak in the
tunneling conductance. The entanglement is completely broken when
$eV>2k_BT_K$, in which a nonentangled Kondo singlet,
$|\Psi\rangle_{ne}=p\!\!\mid\downarrow\uparrow\rangle_{LS}+q\!\!\mid\uparrow\downarrow\rangle_{LS}$,
performs resonant tunneling and yields the side peak when it
reaches the coherent transport channel shown by the dashed line in
Fig.~2 (b). This transport also prohibits backward motion.
Therefore, prohibiting backward motion of coherent spins is a
generic feature of transport in a mesoscopic Kondo system under
bias. This property significantly simplifies the dynamics at
steady-state nonequilibrium. Another crucial feature is the
existence of two coherent transport channels, shown in Fig.~2 (b),
which is attributable to two reservoirs within the coherent
region. Specific proof is given in a previous study~\cite{hong11}
and the basis vectors given later in the text clarify their
existence. Previous studies~\cite{wingreen2,fujii,han,anders} have obtained a similar
result showing Kondo peak splitting with bias. This phenomenon may occur when the two
reservoirs are out of coherence. However, each of the
aforementioned mesoscopic systems should be considered as a
complete coherent system.

Now, we validate the tunneling mechanisms given in Fig.~2 by
obtaining the tunneling conductance that fits the experimental
result. The tunneling current of a mesoscopic system with an
interacting site between two noninteracting reservoirs is given
by~\cite{haug} $$I=(e/\hbar)\int d\omega {\widetilde
\Gamma}(\omega)[f_L(\omega)-f_R(\omega)]\rho^{ss}_{m}(\omega),$$
where $f_{L(R)}(\omega)$ is the Fermi distribution function of the
left (or right) reservoir,
${\widetilde\Gamma}(\omega)=\Gamma^L(\omega)\Gamma^R(\omega)/[\Gamma^L(\omega)+
\Gamma^R(\omega)]$, where $\Gamma^{L(R)}(\omega)=
2\pi\sum_k|V_{km}^{L(R)}|^2\delta(\omega-\omega_k)$ that involves
the reservoir density of states, and $\rho^{ss}_{m}(\omega)$ is
the local density of states (LDOS) at the mediating atom. The
superscript $ss$ means steady-state nonequilibrium. This
simplified form is derived from the well-known Meir-Wingreen
current formula~\cite{meir,hersh} by using the condition of the
proportionate coupling function,
$\Gamma^{L}(\omega)\propto\Gamma^{R}(\omega)$. This proportionate
relation can be applicable to the elastic tunneling shown in
Fig.~2. We employ a constant $\Gamma^{L,R}(\omega)$, indicating a
flat density of states of metallic reservoirs.

Since the electrons in the singlet do not collide with a quasiparticle such as
phonon until the bias excites it, $\rho^{ss}_{m}(\omega)$ in this study must be bias
independent, i.e., $\partial\rho^{ss}_{m}(\omega)/\partial V=0$,
which means one can write the tunneling conductance at zero
temperature as $$dI/dV=(e/\hbar) {\widetilde
\Gamma}(\omega)\rho^{ss}_{m}(\omega)|_{\hbar\omega=eV}.$$
The bias independence of $\rho^{ss}_{m}(\omega)$ will
be retained unless inelastic tunneling is caused by the scattering with
quasiparticles. In other studies on Kondo-involved mesoscopic
systems~\cite{wingreen,schiller,plihal},
$\partial\rho^{ss}_{m}(\omega)/\partial V$ has also been
neglected.

We obtain $\rho_{m\uparrow}^{ss}(\omega)$, which is given by
$\rho_{m\uparrow}^{ss}(\omega)=-(1/\pi){\rm
Im}[G^{+ss}_{mm\uparrow}(\omega)]$, by calculating the on-site
retarded Green's function, $iG^{+ss}_{mm\uparrow}(\omega)=\langle
c_{m\uparrow}|(z+iL)^{-1}|c_{m\uparrow}\rangle$, where
$z=-i\omega+0^+$, $c_{m\uparrow}$ is the fermion operator that
annihilates an up-spin electron at the mediating atom and $L$ is
the Liouville operator defined by $LA={\cal H}A-A{\cal H}$, in
which ${\cal H}$ is the Hamiltonian and $A$ is an operator. To
obtain $G^{+ss}_{mm\uparrow}(\omega)$ using the resolvent form,
one needs a complete set of basis vectors spanning the Liouville
space. We have obtained a complete set of orthonormal basis
vectors~\cite{hong11,hong10} describing $c_{m\uparrow}(t)$ that is
driven by the Hamiltonian of the two-reservoir Anderson impurity
model,
\begin{equation}
{\cal H}={\cal H}_0^L+{\cal
H}_0^R+\sum_{\sigma}\epsilon_mc^\dagger_{m\sigma} c_{m\sigma}
+Un_{m\uparrow}n_{m\downarrow}+{\cal H}_C, \label{hamiltonian}
\end{equation}
where ${\cal
H}_0^{L,R}=\sum_{k,\sigma}(\epsilon_k-\mu^{L,R})c^{\dagger}_{k\sigma}
c_{k\sigma}$, ${\cal
H}_C=\sum_{k,\sigma,\nu=L,R}(V_{km}^{\nu}c^\dagger
_{m\sigma}c_{k\sigma}+V^{\nu*}_{km}c^{\dagger}_{k\sigma}
c_{m\sigma})$, and $\sigma$, $\epsilon_{k}$, $\epsilon_{m}$,
$V_{km}$, $U$, and $\mu$ indicate the electron spin, kinetic
energy, energy level of the mediating atom, hybridization
strength, on-site Coulomb repulsion, and chemical potential,
respectively.

We divide the complete set of basis vectors at
equilibrium~\cite{hong11} into four groups: \\ I:
$\{c_{m\uparrow}, \, \, \, \, n_{m\downarrow}c_{m\uparrow},
 \, \, \, \, j^{\pm L}_{m\downarrow}c_{m\uparrow},  \, \, \, \, j^{\pm R}_{m\downarrow}c_{m\uparrow}\}$,\\
II: $\{c_{k\uparrow}^{L,R}, \, n_{m\downarrow}c_{k\uparrow}^{L,R},
\, j^{\pm L}_{m\downarrow}c_{k\uparrow}^{L}, \,
j^{\pm R}_{m\downarrow}c_{k\uparrow}^{R} \,  |  \, k=1, 2, \cdots, \infty\}$, \\
III: $\{(L_C^nj^{\pm L}_{m\downarrow})c_{m\uparrow},  \, \, \, \,
(L_C^nj^{\pm R}_{m\downarrow})c_{m\uparrow} \, |
 \, \, n=1, \dots, \infty\}$,\\
IV: $\{(L_C^nj^{\pm L}_{m\downarrow})c^{L}_{k\uparrow}$,
$(L_C^nj^{\pm R}_{m\downarrow})c^{R}_{k\uparrow} \, |
 \, \, n, k=1, 2, \cdots, \infty \}$,\\ where $L_C$ denotes the Liouville
operator using ${\cal H}_C$,
$j^+_{m\downarrow}=\sum_k(V_{km}c^\dagger
_{m\downarrow}c_{k\downarrow}+V^*_{km}c^\dagger_{k\downarrow}c_{m\downarrow})$,
and
$j^-_{m\downarrow}=i\sum_k(V_{km}c^\dagger_{m\downarrow}c_{k\downarrow}-
V^*_{km}c^\dagger_{k\downarrow}c_{m\downarrow})$. Groups III and
IV represent multiple trips of a down-spin electron, and they do
not play a role in describing the unidirectional motion of the
Kondo singlet. As a result, the degrees of freedom of the system
are considerably reduced when a bias is applied. The degrees of
freedom are further reduced by neglecting the basis vectors
$n_{m\downarrow}c_{m\uparrow}$ in group I and
$j^{\pm}_{m\downarrow}c_{k\uparrow}$ in group II. The former
describes all higher orders of double occupancy, i.e., from $U$ to
$U^\infty$, because $n_{m\downarrow}^\infty=n_{m\downarrow}$.
Therefore, the basis vector $n_{m\downarrow}c_{m\uparrow}$ must be
neglected in this study. In contrast, the latter basis vectors are
neglected because of their minor contribution to self-energy
compared with $n_{m\downarrow}c_{k\uparrow}$ in group II, whose
members play the role of constructing self-energy. Now, we have a
working Liouville space~\cite{hong11} that is spanned by
$\{\delta j^{+L}_{m\downarrow}c_{m\uparrow}, \, \, \delta
j^{-L}_{m\downarrow}c_{m\uparrow}, \, \, c_{m\uparrow}, \, \,
\delta j^{-R}_{m\downarrow}c_{m\uparrow}, \, \, \delta
j^{+R}_{m\downarrow}c_{m\uparrow}\}  \, \, {\mbox {\rm and}}$
$\{c_{k\uparrow}^L, \, \, \delta
n_{m\downarrow}c_{k\uparrow}^L, \, \, \delta
n_{m\downarrow}c_{k\uparrow}^R, \, \,  c_{k\uparrow}^R \, \, | \, \, k=1, 2, \cdots,
\infty\}.$ We use $\delta$ indicating $\delta A=A-\langle
A\rangle$, where the angular brackets denote the expectation
value, to achieve orthogonality among the basis vectors. For
convenience, we omit the normalization factors $\langle(\delta
j^{\pm L,R}_{m\downarrow})^2\rangle^{1/2}$ and $\langle(\delta
n_{m\downarrow})^2\rangle^{1/2}$ in the denominators of the
corresponding basis vectors.

We construct the matrix ${\rm\bf M}_{\infty\times\infty}\equiv
zI+iL$ in terms of the basis vectors spanning the working
Liouville space. Matrix reduction from ${\rm\bf
M}_{\infty\times\infty}$ to ${\bf M}_{5\times 5}$ is possible for
noninteracting reservoirs~\cite{loewdin}. Thus,
$\rho_{m\uparrow}^{ss}(\omega)$ is given by using the relation
$\rho_{m\uparrow}^{ss}(\omega)=(1/\pi){\rm Re}[({\bf M}^r_{5\times
5})^{-1}]_{33}$, where
\begin{eqnarray} {\rm\bf M}^{r}_{5\times 5}
=\left( \begin{array}{c c c c c} -i\omega' & \gamma_{_{LL}} &
-U^L_{j^-} & \gamma_{_{LR}} & \gamma_{_{j^-}} \\ -\gamma_{_{LL}} &
-i\omega' &
 -U^L_{j^+} & \gamma_{_{j^+}} & \gamma_{_{LR}} \\
U_{j^-}^{L*} &  U_{j^+}^{L*} & -i\omega' &  U^{R*}_{j^+} &
U^{R*}_{j^-} \\  -\gamma_{_{LR}} & -\gamma_{_{j^+}} & -U_{j^+}^R &
-i\omega' & -\gamma_{_{RR}} \\
 -\gamma_{_{j^-}} &  -\gamma_{_{LR}} &
 -U_{j^-}^R  & \gamma_{_{RR}} &  -i\omega'
\end{array} \right). \label{r2r}
\end{eqnarray}
Further, $\omega'\equiv\omega-\epsilon_m-U\langle
n_{m\downarrow}\rangle$ and $\langle n_{m\downarrow}\rangle$
denotes the average number of down-spin electrons occupying the
mediating atom. All the matrix elements, except $U_{j^\pm}^{L,R}$,
have additional self-energy terms
$i\Sigma_{mn}=\beta_{mn}[i\Sigma^L_0(\omega)+i\Sigma^R_0(\omega)]$,
where $\Sigma^{L(R)}_{0}(\omega)=-i\Gamma^{L(R)}/2$ for a flat
wide band. We use $\Delta\equiv(\Gamma^L+\Gamma^R)/4$ as an energy
unit. The coefficients $\beta_{mn}$ are discussed in the following
text.

\begin{figure}
[b] \vspace*{3cm} \includegraphics{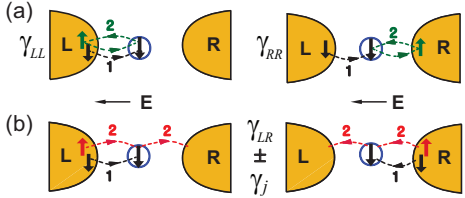} \vspace*{0.0cm} \caption{ (Color
online) Spin dynamics under bias in $\gamma$. The numbers denote
the sequence of coherent motion. A down spin moves into the
mediating atom (blue circle) and performs an exchange [green 2 in
(a)] in $\gamma_{_{LL,RR}}$ and a singlet hopping [red 2 in (b)]
in $\gamma_{_{LR,j}}$. The second part of (b) vanishes because of
the reverse motion.
 }
\end{figure}


The matrix ${\rm\bf M}^{r}$ in Eq. (\ref{r2r}) consists of two
$3\times 3$ blocks that share the central element representing the
mediating atom and two $2\times 2$ blocks at the corners. The
three off-diagonal elements of the $3\times 3$ block represent the
degrees of singlet coupling $\gamma_{_{LL(RR)}}$, e.g.,
\begin{eqnarray*}\gamma_{_{LL}}&=&\frac{\langle \sum_k
i(V_{km}^{*}c_{k\uparrow}^{L}+V_{km}^{*}c_{k\uparrow}^{R})
c^\dagger_{m\uparrow}[j^{-L}_{m\downarrow},j^{+L}_{m\downarrow}]
\rangle}{[\langle (\delta j^{-L}_{m\downarrow})^2\rangle\langle
(\delta j^{+L}_{m\downarrow})^2\rangle]^{-1/2}},
\end{eqnarray*}
and the incoherent double occupancy parameters
\begin{eqnarray*} U_{j^\pm}^{L,R}=\frac{U}{2}[\langle D^{\pm
L,R}_{m\uparrow\downarrow}\rangle+i(1-2\langle
n_{m\downarrow}\rangle)]\frac{\langle j^{\pm
L,R}_{m\downarrow}\rangle}{\langle (\delta j^{\pm
L,R}_{m\downarrow})^2\rangle^{1/2}},
\end{eqnarray*}
where $\langle D^{\pm L,R}_{m\uparrow\downarrow}\rangle=\langle
[n_{m\downarrow},j^{\pm L,R}_{m\downarrow}](1-2
n_{m\uparrow})\rangle/\langle j^{\pm L,R}_{m\downarrow}\rangle$.
Hence, the real part ${\rm Re}[U_{j^\pm}^{L,R}]$ represents the
probability of double occupancy by $j^{+}_{m\downarrow}$ or
$j^{-}_{m\downarrow}$ coming from the left or right reservoir. In
contrast, the elements of the $2\times 2$ corner blocks, i.e.,
\begin{eqnarray*}\gamma_{_{LR}}=\frac{\langle\sum_ki
(V_{km}^{*}c_{k\uparrow}^L+V_{km}^{*}c_{k\uparrow}^R)c^\dagger_{m\uparrow}
[j^{-L}_{m\downarrow},j^{+R}_{m\downarrow}]\rangle}{[\langle
(\delta j^{-L}_{m\downarrow})^2\rangle\langle (\delta
j^{+R}_{m\downarrow})^2\rangle]^{-1/2}},\end{eqnarray*} and
$\gamma_{_{j^\mp}}=\gamma_{_{j}}$, where
\begin{eqnarray*}\gamma_{_{j^\mp}}=\frac{\langle\sum_ki
(V_{km}^{*}c_{k\uparrow}^L+V_{km}^{*}c_{k\uparrow}^R)
c^\dagger_{m\uparrow}[j^{\mp L}_{m\downarrow},j^{\mp
R}_{m\downarrow}]\rangle}{[\langle (\delta j^{\mp
L}_{m\downarrow})^2\rangle\langle (\delta j^{\mp
R}_{m\downarrow})^2\rangle]^{-1/2}}\end{eqnarray*} describe the
transition between two reservoirs. In Fig.~3, we present graphical
illustrations of the third order hybridization processes embedded
in $\gamma$ under bias. Singlet partner change and singlet hopping
will occur to perform the process of Fig.~2 (a). In contrast, only
singlet hopping is needed for Fig.~2 (b). The basis vectors in
groups III and IV cause incoherent or backward motion that must be
excluded in describing the unidirectional coherent motion at
steady-state nonequilibrium. Therefore, neglecting groups III and
IV is legitimate. The unidirectional movement of the Kondo singlet
discussed in Fig.~2 guarantees $\gamma_{_{LR}}=\gamma_{_{j}}$ in
Fig.~3 (b). This equality is considered as a condition of
steady-state nonequilibrium.

The $5\times 5$ matrix of Eq. (\ref{r2r}) gives three coherent and
two incoherent poles in the LDOS. One of three coherent poles is
located at the Fermi level and the other two are at
$\pm\hbar\omega_{rt\ell}$, which are the levels represented by the
dashed lines in Fig.~2 (b). The subscript $rt\ell$ indicates the
resonant tunneling level. Hence, the tunneling current rapidly
increases when the bias voltage reaches $\pm\hbar\omega_{rt\ell}$.
A simple atomic limit analysis using the same ${\rm
Re}[U^{L,R}_{j^\pm}]$ gives
$\hbar\omega_{rt\ell}=\pm[(\gamma_{_{LL}}^2+\gamma_{_{RR}}^2)/2]^{1/2}+O(U^{-2})$
and the spectral weight of the zero-bias peak as
$Z_F=\gamma_{_{LL}}^2\gamma_{_{RR}}^2/2{\rm
Re}[U^{L,R}_{j^\pm}]^2(\gamma_{_{LL}}^2+\gamma_{_{RR}}^2)+O(U^{-2})$.
The latter expression shows that the zero-bias peak is suppressed
when $\gamma_{_{LL}}$ and $\gamma_{_{RR}}$ are imbalanced.

We show in the following that ${\rm Re}[U^{L,R}_{j^+}]>{\rm
Re}[U^{L,R}_{j^-}]$ is introduced for symmetric reservoirs, which
means that operator $j_{m\downarrow}^+$ induces more double
occupancy than $j_{m\downarrow}^-$. For asymmetric reservoirs such
as substrate ($L$) and tip ($R$) in scanning tunneling
spectroscopy (STS), we use ${\rm Re}[U^{L}_{j^+}]>{\rm
Re}[U^{L,R}_{j^-}]>{\rm Re}[U^{R}_{j^+}]$ to reflect different
properties of reservoirs. In contrast, we adopt ${\rm
Re}[U^{L}_{j^-}]={\rm Re}[U^{R}_{j^-}]$ because of the
steady-state property of the current operator $j_{m\downarrow}^-$.

Before obtaining the $dI/dV$ line shapes, we first determine the
coefficients $\beta_{mn}$. ${\rm Re}[\beta_{25}]$, for example, is
given by ${\rm Re}[\beta_{25}]=\{\langle j^{-L}_{m\downarrow}(1-2
n_{m\uparrow})\rangle\langle j^{+R}_{m\downarrow}(1-2
n_{m\uparrow})\rangle+(1-2\langle n_{m\downarrow}\rangle)^2\langle
j^{+L}_{m\downarrow} \rangle\langle
j^{-R}_{m\downarrow}\rangle\}/4\langle(\delta
n_{m\downarrow})^2\rangle\sqrt{\langle(\delta
j^{+L}_{m\downarrow})^2\rangle} \sqrt{\langle(\delta
j^{-R}_{m\downarrow})^2\rangle}$. The operators in the first term
describe the self-energy dynamics that avoids double
occupancy~\cite{hong11}. We assume the same contribution of
self-energy dynamics to all ${\rm Re}[\beta_{mn}]$ and the same
relative fluctuations. Therefore, the differences among ${\rm
Re}[\beta_{mn}]$ are attributable to the different signs for the
current, i.e., $\langle j^{-L}_{m\downarrow}\rangle=-\langle
j^{-R}_{m\downarrow}\rangle<0$. From the property  
$\langle j^{+L,R}_{m\downarrow}\rangle>0$, symmetric ${\rm Re}[\beta_{mn}]$
have the following mutual relations: ${\rm
Re}[\beta_{12}]={\rm Re}[\beta_{14}]={\rm Re}[\beta_{15}]<{\rm Re}[\beta_{11}]
={\rm Re}[\beta_{22}]={\rm Re}[\beta_{44}]={\rm
Re}[\beta_{55}]={\rm Re}[\beta_{24}]={\rm Re}[\beta_{25}]={\rm
Re}[\beta_{45}]$ and ${\rm Re}[\beta_{33}]=1$. In this study, we
choose ${\rm Re}[\beta_{11}]=0.255$ and ${\rm
Re}[\beta_{14}]=0.245$ based on the standard value ${\rm
Re}[\beta_{mn}]=0.25$ that is obtained at the atomic
limit~\cite{hong11}. The same relative fluctuations give vanishing
${\rm Im}\beta_{mn}$ because they are given by the difference in
the relative fluctuations.

The experimental $dI/dV$ line shapes under consideration are those
of a quantum point contact with the closest side peaks given in
Fig.~1 (b) of Ref.~\cite{sarkozy} and the STS for a Co atom placed
on a Cu$_2$N layer on a Cu (100) substrate of Ref.~\cite{otte}.
For the former, we employ the scenario of spontaneous formation of
a localized spin at the bound state~\cite{rejec,ihna}. Therefore,
the Hamiltonian of Eq. (\ref{hamiltonian}) is applicable to both
cases. The gate voltage dependence in the former system will be
described in a future study. The theoretical $dI/dV$ line shapes
given in Fig.~4 are obtained by using the matrix elements given in
Table I. We set the energy unit to $\Delta\approx$ 1.5 and 5 meV
for Figs.~4 (a) and 4 (b), respectively. The fittings are
remarkably good.

\begin{figure}
[t] \vspace*{3.5cm} \includegraphics{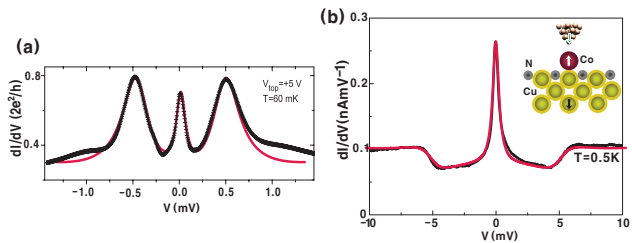} \vspace*{0.0cm}
\caption{(Color online) Comparison of the theoretical line shape
(red) using $\Delta\approx 1.5$ meV (a) and $\Delta\approx 5$ meV
(b) with the experimental data (black) of the closest side peaks
reported in Ref.~\cite{sarkozy} and the STS line shape of
Ref.~\cite{otte}, respectively. We choose
$\widetilde{\Gamma}/\Delta=0.82$ in (a) and an arbitrary unit for
the theoretical $dI/dV$ in (b). Inset is STS
setup~\cite{otte,choi}. }
\end{figure}


\begin{table}[t]
Table I: \textbf{Matrix elements for Fig. 4}

\begin{tabular}{c c c c c c c c}
\hline\hline  \, \, & \, $\gamma_{_{LL}}$ \, & $\gamma_{_{RR}}$ \,
&
 $\gamma_{_{j,LR}}$ \, & $
{\rm Re}U^{L,R}_{j^-}$ & ${\rm Re}U_{j^+}^{L}$ & ${\rm
Re}U_{j^+}^{R}$ & ${\rm Im}U_{j^\pm}^{L,R}$
\\ [0.5ex] \hline \\
(a) & 0.38  & 0.38 &  0.5 & 0.98 & 1.1 & 1.1  & 0  \\
(b) & 0.86  & 0.76 &  0.43 & 2.8 & 7.0 & 1.62  & 0  \\
[1ex] \hline
\end{tabular} \label{table1}
\end{table}


As listed in Table I, we adopt a symmetric Kondo coupling
($\gamma_{_{LL}}=\gamma_{_{RR}}$) and ${\rm Re}[U_{j^+}^{L}]={\rm
Re}[U_{j^+}^{R}]>{\rm Re}[U_{j^-}^{L,R}]$ in Fig.~4 (a) and an
asymmetric Kondo coupling ($\gamma_{_{LL}}\neq\gamma_{_{RR}}$) and
${\rm Re}[U_{j^+}^{L}]>{\rm Re}[U_{j^-}^{L,R}]>{\rm
Re}[U_{j^+}^{R}]$ in Fig.~4 (b). The zero-bias peak in Fig.~4 (a)
is slightly suppressed by different contributions to double
occupancy by operators $j^+$ and $j^-$ and the position of the
side peak matches $\hbar\omega_{rt\ell}$ pretty well. The
deviation outside the side peaks indicates that the range of bias
independence of the LDOS covers the two side peaks. It is
noteworthy that the line shape of Fig.~4 (b) shows a dip at zero
bias when the insulating layer is removed~\cite{mano}. This
implies that a strong Kondo coupling is established by inserting
the insulating layer. Choi {\it et al}.~\cite{choi}, who studied
the same system, observed that the meaningful structure of the
line shape disappears when a Co atom is placed on top of a N atom.
This indicates that the N atoms surrounding a Cu atom in a Cu$_2$N
layer play the role of barrier that suppresses fluctuations and
enhance ${\rm Re}[U_{j^+}^{L}]$ and the axial Kondo coupling
connecting tip, Co atom, and Cu substrate. The large values of
$\gamma_{_{LL}}$ and $\gamma_{_{RR}}$ given in Table I verify this
fact. Choi {\it et al}.~\cite{choi} also show that the shoulder in
Fig.~4 (b) is a variation of a coherent side peak.

In conclusion, our theoretical study clarifies that there are two
different transport channels (Fig.~2). One uses the entangled
Kondo singlet that connects the Kondo clouds in both reservoirs.
Transport by the entangled Kondo singlet forms the zero-bias peak.
The other uses resonant tunneling of a nonentangled Kondo singlet
through the coherent tunneling level. This tunneling mechanism
forms the side peak. Comparisons of the theoretical $dI/dV$ line
shapes with those of the experimental ones (Fig.~4) clearly
demonstrate the existence of the two transport channels.

The author thanks P. Coleman for suggesting the entangled singlet and
A. Millis, N. Andrei, J. E. Han, E. Yuzbashyan, P. Kim, S.-W. Cheong,
and P. Fulde for valuable discussions.
This research was supported by the Basic Science Research Program through the NRF, Korea
(2012R1A1A2005220), and was partially supported by a KIAS grant
funded by MEST.






\begin{thebibliography}{10}

\bibitem {wingreen2} N. S. Wingreen and Y. Meir, Phys. Rev. B {\bf
49}, 11040 (1994).

\bibitem{fujii} T. Fujii and K. Ueda, Phys. Rev. B {\bf 68}, 155310 (2003).

\bibitem{han} J. E. Han and R. J. Heary, Phys. Rev. Lett. {\bf
99}, 236808 (2007).

\bibitem{anders} F. B. Anders, Phys. Rev. Lett. {\bf
101}, 066804 (2008).

\bibitem {wilson} H. R. Krishna-murthy, J. W. Wilkins, and K. G. Wilson,
 Phys. Rev. B {\bf 21}, 1003 and 1044 (1980).

\bibitem{sarkozy}  S. Sarkozy \emph{et al.}, Phys. Rev. B {\bf
79}, 161307(R) (2009).

\bibitem{otte}  A. F. Otte \emph{et al}. Nat. Phys. {\bf 4}, 847 (2008).

\bibitem{hong11} J. Hong, Phys. J. Phys. Condens. Matter {\bf 23}, 275602 (2011).

\bibitem{haug}  H. Haug and A.-P. Jauho, {\it Quantum Kinetics in Transport and Optics of
Semiconductors} (Springer-Verlag, Berlin, 1996) Chap. 12.

\bibitem{meir} Y.  Meir and N. S. Wingreen, Phys. Rev. Lett. {\bf 68}, 2512
(1992).

\bibitem{hersh} S. Hershfield, J. H. Davies, and J. W. Wilkins, Phys. Rev. B {\bf 46}, 7046
(1992).

\bibitem{wingreen} V. Madhavan, W. Chen, T. Jamneala, M. F. Crommie, and N. S. Wingreen, Phys. Rev. B {\bf
64}, 165412 (2001).

\bibitem{schiller} A. Schiller and S. Hershfield, Phys. Rev. B {\bf
61}, 9036(2000).

\bibitem{plihal} M. Plihal and J. W. Gadzuk, Phys. Rev. B {\bf 63},
085404 (2001).

\bibitem{hong10} J. Hong, J. Phys.: Condens. Matter {\bf
23}, 225601 (2011).

\bibitem{loewdin} P. O. L\"{o}wdin, J. Math. Phys. {\bf 3}, 969 (1962).

\bibitem{mano} H. C. Manoharan, C. P. Lutz, and D. M. Eigler, Nature
{\bf 403}, 512 (2000).

\bibitem{choi} T. Choi, C. D. Ruggiero, and J. A. Gupta, J. Vac. Sci. Technol.
B {\bf 27}, 887 (2009).

\bibitem{rejec} T. Rejec and Y. Meir, Nature {\bf 442}, 900
(2006).

\bibitem{ihna} S. Ihnatsenka and I. V. Zozoulenko,
Phys. Rev. B {\bf 76}, 045338 (2007).

\end{thebibliography}
\end{document}